# STOCHASTIC DIFFERENTIAL GAMES IN A NON-MARKOVIAN SETTING *

ERHAN BAYRAKTAR † AND H. VINCENT POOR ‡

**Abstract.** Stochastic differential games are considered in a non-Markovian setting. Typically, in stochastic differential games the modulating process of the diffusion equation describing the state flow is taken to be Markovian. Then Nash equilibria or other types of solution such as Pareto equilibria are constructed using Hamilton-Jacobi-Bellman (HJB) equations. But in a non-Markovian setting the HJB method is not applicable. To examine the non-Markovian case, this paper considers the situation in which the modulating process is a fractional Brownian motion. Fractional noise calculus is used for such models to find the Nash equilibria explicitly. Although fractional Brownian motion is taken as the modulating process because of its versatility in modeling in the fields of finance and networks, the approach in this paper has the merit of being applicable to more general Gaussian stochastic differential games with only slight conceptual modifications. This work has applications in finance to stock price modeling which incorporates the effect of institutional investors, and to stochastic differential portfolio games in markets in which the stock prices follow diffusions modulated with fractional Brownian motion.

**1. Introduction.** The study of stochastic differential games with controls is a part of game theory that is relatively unknown, even though it has significant potential for application as noted by Øksendal and Reikvam [29]. Prior work in this area has focused on the examination of such games in a Markovian setting (see below). In this paper we will study a type of non-Markovian stochastic differential game. In particular, we will consider a game in which the one-dimensional state $X_t$ follows the following stochastic differential equation:

$$dX_t = \mu(t, X_t, v_1, ..., v_N)dt + \sigma(t, X_t, v_1, ..., v_N)dB_t^{(H)}, \tag{1.1}$$

$$\mu : [0, T] \times \mathbb{R} \times \Upsilon_1 \times ... \times \Upsilon_N \to \mathbb{R}, \tag{1.2}$$

$$\sigma : [0, T] \times \mathbb{R} \times \Upsilon_1 \times ... \times \Upsilon_N \to \mathbb{R}, \tag{1.3}$$

where $v_i \in \Upsilon_i \subset \mathbb{R}^{\nu_i}$ is the control of the $i$th player over the state and is adapted to the natural filtration of $B^H$. Here $T$ is the expiration date of the game, and $B^H$ is a fractional Brownian motion (fBm) with Hurst parameter $H \in (\frac{1}{2}, 1)$. $B^H$ is defined as an almost surely (a.s.) continuous zero mean Gaussian process having the autocorrelation structure given by

$$E\left\{B_t^H B_s^H\right\} = \frac{1}{2}\left\{|s|^{2H} + |t|^{2H} - |t-s|^{2H}\right\} \tag{1.4}$$

---

*This research was supported by the U.S. Office of Naval Research under Grant No. N00014-03-1-0102.

†Department of Mathematics, University of Michigan, 525 East University, Ann Arbor, MI 48109 erhan@umich.edu, tel: 734-936-9973, fax: 734-763-0937 .

‡Department of Electrical Engineering, Princeton University, Princeton, NJ 08544, poor@princeton.edu, tel: 609-258-3500, fax: 609-258-3745.

Each agent wants to maximize its own pay-off (with this feature the problem differs from the usual optimal control problem):

$$J_i(x) = E^x \left\{ \int_0^T f_i(t, X_t, v_t)dt + K_i(T, X_T) \right\}, \tag{1.5}$$

where $E^x\{\cdot\}$ denotes conditional expectation given $X_0 = x$. In this paper we consider only the case in which the state and the source of randomness are one-dimensional. The results can be extended to the case in which there are multiple sources of randomness and multiple controlled states (see [4]).

Typically in this type of setting the modulating process in (1.1) is taken to be Brownian motion, i.e., $H = 1/2$ and the controls of the players are Markovian. Then Nash equilibria or other types of solution such as Pareto equilibria are constructed using the Hamilton-Jacobi-Bellman (HJB) equations. (See e.g. Friedman [10], Gaidov [12], [13], [11] Nilakantan [25], Øksendal and Reikvam [29], and Pravin [34].) However, fBm is not a Markov process for any $H$ other than $\frac{1}{2}$, and therefore this approach does not work for the general case of (1.1). Here we will develop a quasi-martingale approach to the solution of this problem using the fractional noise calculus developed by Duncan et al. [8], and Øksendal and Hu [17] which generalizes white noise calculus (see [20]) to develop an integration theory with respect to fBm. The key to our solution will be the fractional Clark-Ocone formula developed by Øksendal and Hu [17]. The integrals in (1.1) are Wick type integrals (see Definition 3.3) rather than Stieltjes integrals (defined pathwise; see e.g. [35]). The motivation for using Wick type integrals is as follows: The pathwise integral $\int_0^t f_s \delta B_s^H$ with respect to fBm does not in general have zero mean, i.e. $E\left\{\int_0^t f_s \delta B_s^H\right\} \neq 0$. However the Wick type integral $\int_0^t f_s dB_s^H$ has zero mean, i.e. $E\left\{\int_0^t f_s dB_s^H\right\} = 0$. Therefore in a stochastic differential equation (SDE) of the form

$$dX_t = b(X_t)dt + \sigma(X_t)dB_t^H, \tag{1.6}$$

the volatility term $\sigma(X_t)dB_t^H$ does not contribute to the mean rate of change, as it does in SDE's defined by pathwise integrals. Since seperating the random fluctuations from the mean rate of change is desirable for our purposes, we prefer to use Wick type integrals for defining the integrals with respect to fBm (see [8]). (Note also that only in the Wick type calculus are the standard tools of Itô calculus, such as an Itô representation theorem, available.) See [6] for applications of Wick calculus to pricing weather derivatives, and [9] and [17] for further applications of Wick type calculus particularly in finance.

Fractional noise calculus reduces to white noise calculus when $H$ is replaced by $1/2$. Moreover the integrals of adapted processes in this framework are equal to the Itô integrals of these procesess with respect to Brownian motion. Hence our results hold in particular for the standard framework, i.e. when the modulator is a Brownian motion, and the integrals in (1.1) are taken to be Itô integrals.

This work has an immediate application in finance, to stock price modeling when long-range dependence is accounted for in the model. The stock prices are considered to be states in this setting while the agents are *not price takers*, *i.e.* their trading change the price level. This models a market with institutional investors who make large transactions and therefore influence prices. These investors find themselves in a random environment due to the existence of small investors. The small investors are typically inert, that is they do not trade for long time intervals. A micro-structure model taking the inertness of the agents into account is constructed in [2]. It is shown that the prices arising from the interaction of the small agents can be approximated by geometric fractional Brownian motion. The game theoretic setting in this paper is an extension to the results of [2] in the sense that, we start by assuming that the noise in the environment can be modeled by an fBm differential in the controlled stochastic differential equation (1.1), which models the noise due to the trades of the small investors.

Another possible application is stochastic differential portfolio games, which are studied by Browne in a Brownian motion setting in [7]. This formulation is applicable to the analysis of traders who are competing for a bonus, or to fund managers whose funds are invested in different markets, and who achieve rewards based on the relative performance of their funds. Yet another possible application is in stochastic goodwill problems (finding the optimal advertising policy for the maximization of product image) in advertising when there is more than one good of the same kind in competition. (See [24] for stochastic goodwill problems in a stochastic optimal control setting.)

By adapting the fractional noise machinery, our results will hold for more general Gaussian modulators in the state flow dynamics.

However we state the results in terms of fBm to emphasize the fact that the game under consideration becomes non-Markovian, and also because this case admits an explicit equilibrium. Another motivation for this model is the fact that fBm is frequently used for modeling in various areas of research (see [3], [5] for applications in finance and [1], [33], [26], [27], [28] for applications other than finance).

The rest of the paper is organized as follows. In Section 2 we give a Nash-equilibrium theorem. In Section 3 we introduce the necessary tools from the fractional noise calculus that we use in the proof of the equilibrium theorem in Section 4. Finally in Section 5 we give a sketch of how to extend the fractional noise machinery to more general Gaussian modulation processes.

**2. Nash Equilibrium in a Linear Game of $N$ players.** For ease of exposition we consider first a one dimensional state equation, with the drift and diffusion coefficients controlled linearly by the players:

$$dX_t = rX_t dt + \sum_{i=1}^{N} \alpha_i(t)u_i(t)dt + C\sum_{i=1}^{N} \beta_i(t)v_i(t)dt + \sum_{i=1}^{N} \beta_i(t)v_i(t)dB^H(t), \quad (2.1)$$

where $B^H$ denotes the one-dimensional version of $B^{(H)}$ with $H_1 = H$. The initial state will be denoted by $X_0 = x$. The pay-off function of player $i$ will be of the form:

$$J_i(x) = E_\mu^x \left\{ \int_0^T \frac{c_i u_i^{\gamma_i}(t)}{\gamma_i} dt + \frac{b_i X_T^{\gamma_i'}}{\gamma_i'} \right\}; \quad (2.2)$$

that is, players are constant relative risk averse (CRRA). Here $\mu$ is the measure on the sample space under which the canonical process is an fBm. Player $i$ controls the state by its choice of actions $(u_i, v_i)$. We assume that $\alpha_i : [0, T] \to \mathbb{R}$ is bounded for each $i \in \{1, ..., N\}$. The coefficients functions $\beta_i : [0, T] \to \mathbb{R}$ will appear in the definition of admissible strategies.

Since this game is in a non-Markovian setting, it cannot be solved via the HJB method. Instead, as noted in Section 1, we will employ the recently developed fractional Wick calculus, which we describe briefly in Section 3 and the fractional Clark-Ocone formula (which is given along with the proof of the equilibrium theorem) to find Nash equilibria for this game. Observe that $u_i$ affects the drift of the state and also appears in the pay-off function. It can be interpreted as a cost for the player, *i.e.* for gaining a certain amount of riskless increase the player pays an associated cost. Whereas by choice of $v_i$ the player does not have to pay a cost for an associated gain (since this action does not appear in the pay-off function), but it must take some risk (since $v_i$ affects the diffusion coefficient in addition to the drift).

Let us introduce the following notation which is necessary to define the admissible strategies and for the statement of the theorem.

Define $K$ as

$$K(t) = \frac{C(Tt - t^2)^{\frac{1}{2} - H}}{2H(2H-1)\Gamma(2H-1)\Gamma(2-2H)\cos(\pi(H - \frac{1}{2}))}, \quad \text{for } t \in [0, T], \quad (2.3)$$

and define $\zeta$ by

$$\begin{aligned}((-\Delta)^{-(H-1/2)}\zeta_t)(s) &= ((-\Delta)^{-(H-1/2)}K)(s), \quad 0 \leq s \leq t, \\ \zeta_t(s) &= 0 \quad s < 0 \quad \text{or} \quad s > t,\end{aligned} \quad (2.4)$$

where the operator $(-\Delta)^{-(H-1/2)}$ operates on a test function $f$ as

$$((-\Delta)^{-(H-1/2)}f)(x) = \frac{1}{2\Gamma(2H-1)\cos(\pi(H-1/2))} \int_{-\infty}^{\infty} |x-t|^{2H-2} f(t) dt. \quad (2.5)$$

where $\Gamma$ is the gamma function and is given by $\Gamma(x) = \int_0^\infty t^{x-1} e^{-t} dt$ for $x > 0$. The existence of such $\zeta$ is guaranteed by [16].

Define $\hat{\mu}$ and $\eta$ by

$$\eta(T) := \frac{d\hat{\mu}}{d\mu} := \exp\left(-\int_0^T K(s) dB_s^H - \frac{1}{2}|K|_\phi^2\right), \quad (2.6)$$

where $|K|_\phi^2$ is given by $|K|_\phi^2 = \int_{\mathbb{R}_+^2} K(s) K(t) \phi(s,t) ds dt$, and where

$$\phi(s,t) = H(2H-1)|s-t|^{2H-2}; \quad s, t \in \mathbb{R}_+. \quad (2.7)$$

Here $\mu$ denotes the probability measure under which $B^H$ is an fBm with Hurst parameter $H$. Note that integrals of deterministic functions with respect to (w.r.t) fBm are well defined, as will become clear in Section 3.

And finally define $\rho$ by $\rho(t, w) := E_\mu\{\eta(T) | \mathcal{F}_t\}$, where $\mathcal{F}_t$ is the $\sigma$-algebra generated by $\{B_s^H, s \leq t\}$.

Now we will introduce the solution concept of Nash equilibrium in our context. We consider a set $\mathcal{A} = \mathcal{A}_1 \times ... \times \mathcal{A}_N$ of admissible strategies for which the admissibility conditions are adaptedness w.r.t. the filtration generated by fBm and the following integrability condition

$$\beta_i v_i \in \mathcal{L}_\phi^{1,2}(\hat{\mu}), \quad (2.8)$$

where $\mathcal{L}_\phi^{1,2}(\hat{\mu})$ denotes the completion of the set of all $\mathcal{F}_t$ adapted processes $f$ such that

$$\| f \|_{\mathcal{L}_\phi^{1,2}(\hat{\mu})} := E_{\hat{\mu}}\left\{\int_{\mathbb{R}} \int_{\mathbb{R}} f(s) f(t) \phi(s,t) ds dt\right\} + E_{\hat{\mu}}\left\{\left(\int_{\mathbb{R}} D_s^\phi f(s) ds\right)^2\right\} < \infty. \quad (2.9)$$

Here $D_s^\phi F = \int_{\mathbb{R}} \phi(s,t) D_t F dt$, where $D_s F$ denotes the Hida-Malliavin derivative of $F$, which will be introduced in Section 3.

DEFINITION 2.1. *The strategy $z^e = (u^e, v^e) \in \mathcal{A}$ is called a Nash-equilibrium strategy if, for each $i$, player $i$'s action $z_i^e = (u_i, v_i) \in \mathcal{A}_i$ is a best response to its opponents,* i.e.

$$J_i^x(z_1^e, ..., z_{i-1}^e, z_i, z_{i+1}^e, ..., z_N^e) \leq J_i^x(z^e), \tag{2.10}$$

*for all $x$, for each player $i$ and for all $z_i \in \mathcal{A}_i$.*

Then we have the following Nash equilibrium theorem:

THEOREM 2.2. *Consider the game given by (2.1) and (2.2). Then the following conditions ((2.11), (2.12)) are necessary and sufficient for the existence of a Nash-equilibrium:*

$$\gamma_i' = \gamma' \quad \text{for } i = 1, ..., N, \quad \text{and} \tag{2.11}$$

$$\begin{aligned}
\sum_{i=1}^{N} \int_0^T & m^{\frac{1}{\gamma_i-1}} b_i^{\frac{1}{\gamma_i-1}} \alpha_i(t)^{\frac{\gamma_i}{\gamma_i-1}} e^{-rt \frac{\gamma_i}{\gamma_i-1}} \exp\left(\frac{\gamma_i}{2(1-\gamma_i)^2}|\zeta_t|_\phi\right) dt \\
& + m^{\frac{1}{\gamma'-1}} e^{-rT \frac{\gamma'}{\gamma'-1}} \exp\left(\frac{2\gamma'|K|_\phi}{2(1-\gamma')^2}\right) = x,
\end{aligned} \tag{2.12}$$

*has a solution $m^* \in \mathbb{R}$.*

*Let $\left((u_1^e, v_1^e), ...., (u_N^e, v_N^e)\right)$ denote the agents' Nash-equilibrium strategies. The first components of the equilibrium strategies are uniquely determined by*

$$u_i^e(t) = \left(\frac{m^* b_i \alpha_i(t)}{c_i} e^{-rt} \rho(t,w)\right)^{\frac{1}{\gamma_i-1}}, \quad \text{for } i = 1, ..., N, \tag{2.13}$$

*while the second component of the players' strategies will be any adapted (to the filtration of fBm) processes satisfying the following constraint:*

$$\begin{aligned}
e^{-rt} \sum_{i=1}^{N} \beta_i v_i^e(t) = & (m^*)^{\frac{1}{\gamma'-1}} \frac{K(t)}{1-\gamma'} \exp\left(\frac{1}{1-\gamma'} \int_0^t K(s) dB_s^H - \frac{C}{1-\gamma'} \int_t^T K(s) ds \right. \\
& \left. + \frac{2-\gamma}{2(1-\gamma^2)}|K|_\phi^2 - \frac{1}{1-\gamma}|K 1_{[0,t]}|_\phi^2 - rT \frac{\gamma'}{\gamma'-1}\right) - \int_0^T \sum_{i=1}^{N} \alpha_i(u)^{\frac{\gamma_i}{\gamma_i-1}} \left(\frac{m^* b_i}{c_i}\right)^{\frac{1}{\gamma_i-1}} \\
& \times \frac{\zeta_u(t)}{1-\gamma_i} e^{-ru \frac{\gamma_i}{\gamma_i-1}} \exp\left(\frac{1}{1-\gamma_i} \int_0^t \zeta_u(s) dB_s^H - \frac{C}{1-\gamma_i} \int_t^u \zeta_u(s) ds \right. \\
& \left. + \frac{2-\gamma_i}{2(1-\gamma_i)^2}|\zeta_u|_\phi^2 - \frac{1}{1-\gamma_i}|\zeta_u 1_{[0,t]}|_\phi^2\right) du,
\end{aligned} \tag{2.14}$$

*where 1 stands for the indicator function. Finally, the state at time $T$ at Nash equilibrium is given by*

$$F^e = (m^*)^{\frac{1}{\gamma'-1}} \eta(T)^{\frac{1}{\gamma'-1}} e^{\frac{-rT}{\gamma'-1}}. \tag{2.15}$$

These results can be extended to games with multiple numbers of controlled states and multiple sources of randomness (see [4]).

Before proving Theorem 2.2 we will, in the next section, give a brief review of basic results from fractional noise calculus, mostly following the treatment by Hu and Øksendal [17]. (An extended version of the following section can be found in [4].) [1]

**3. Fractional Noise Calculus.** We will start this section by introducing the necessary ingredients for the definition of the the stochastic integral in (1.1). In what follows $L^2_\phi(\mathbb{R})$ will denote the completion of the set of measurable functions satisfying

$$|f|^2_\phi := \int_{\mathbb{R}^2} f(s)f(t)\phi(s,t)dsdt < \infty. \tag{3.1}$$

*Remark*: It is shown by Taqqu and Pipiras [30] that the set of functions satisfying (3.1) is not a complete space.

The stochastic integrals of deterministic functions in $L^2_\phi(\mathbb{R})$ w.r.t. fBm are well defined (see [14]). For $f \in L^2_\phi(\mathbb{R})$, we will denote its integral w.r.t fBm by $<w, f> := \int_{\mathbb{R}} f(t)dB^H_t$.

The probability space in our game will be $\Omega = \mathcal{S}'(R)$, the space of tempered distributions (the dual space of $\mathcal{S}(\mathbb{R})$, the Schwarz space of rapidly decreasing functions) equipped with the weak-star topology. And we will take the events to be Borel subsets of $\mathcal{S}'(R)$. By the Bochner-Minlos theorem there exists a probability measure $\mu$ on $\Omega$ such that, $<\cdot, f>: \Omega \to \mathbb{R}$ is a Gaussian random variable with mean 0, and variance $|f|^2_\phi$ (see [17]).

We will now introduce the Wiener chaos expansion of random variables in $L^2(\mu)$. We first must find the orthonormal basis for $L^2_\phi(\mathbb{R})$. Recall that the Hermite functions (see e.g. Appendix C of [15]), which we will denote by $(z_n)$, form an orthonormal basis for $L^2(\mathbb{R})$.

Let us define the map from the space of functions satisfying (3.1) into $L^2(\mathbb{R})$ by

$$(I_\phi f)(u) = c_H \int_u^\infty (t-u)^{H-\frac{3}{2}} f(t)dt, \tag{3.2}$$

where $c_H = \sqrt{\frac{H(2H-1)\Gamma(\frac{3}{2}-H)}{\Gamma(H-\frac{1}{2})\Gamma(2-2H)}}$ (here, as before, $\Gamma$ denotes the gamma function). This map preserves the inner product, and the Hermite functions are in the range of this map. Let $I_\phi^{-1}$ denote the inverse map of $I_\phi$. (For summable functions this inverse exists and is proportional to the Liouville differential of order $H - \frac{1}{2}$ [32], since $I_\phi(f)$ is proportional to the fractional integral of $f$ of order $H - \frac{1}{2}$.) Now we see that the set $(e_n = I_\phi^{-1}(z_n))_{n=1,2,...}$ constitutes an orthonormal basis for $L^2_\phi(\mathbb{R})$.

Let $\mathcal{J}$ denote the set of all (finite) multi-indices of non-negative integers. Then for $\alpha = (\alpha_1, \alpha_2, ..., \alpha_m) \in \mathcal{J}$, define

$$H_\alpha(w) := h_{\alpha_1}(<w, e_1>)...h_{\alpha_m}(<w, e_m>). \tag{3.3}$$

Note that $E_\mu\{H_\alpha H_\beta\} = 0$ if $\alpha \neq \beta$, and $E_\mu\{H^2_\alpha\} = \alpha!$. Now we can state what is known as the chaos decomposition for the elements of $L^2(\mu)$ (see [17]): Every $F \in L^2(\mu)$ can be decomposed uniquely as $F(w) = \sum_{\alpha \in \mathcal{J}} c_\alpha H_\alpha(w)$ where $c_\alpha \in \mathbb{R}$ for all $\alpha \in \mathcal{J}$.

---

[1] Although the original name given by Hu and Øksendal in [17] to this kind of calculus was fractional white noise calculus, we prefer to omit 'white' from the name since white suggests independence at each point in time, and here the noise considered does not have this property.

For defining the integration w.r.t fBm of random functions we will make use of the Hida test function space (a subspace of $L^2(\mu)$), and Hida distribution space (a superset of $L^2(\mu)$) which we denote by $(\mathcal{S})_H$ and $(\mathcal{S})_H^*$ respectively. (See [36] for the definitions of these two spaces.) Let $\psi(w) = \sum_{\alpha \in \mathcal{J}} a_\alpha H_\alpha(w) \in (\mathcal{S})_H$ and $G(w) = \sum_{\beta \in \mathcal{J}} b_\beta H_\beta(w)$, and denote the action of $G$ on $\psi$ by $<< G, \psi >> := \sum_{\alpha \in \mathcal{J}} \alpha! a_\alpha b_\alpha$.

For defining the integral w.r.t fBm it is necessary to define $(\mathcal{S})_H^*$-valued Pettis integrals as follows.

DEFINITION 3.1. *A function $Z : \mathbb{R} \to (\mathcal{S})_H^*$ is $(\mathcal{S})_H^*$ integrable if $<< Z(t), \psi >> \in L^1(\mathbb{R})$ for all $\psi \in (\mathcal{S})_H$. In this case, the $(\mathcal{S})_H^*$ integral of $Z$, denoted by $\int_\mathbb{R} Z(t) dt$, is the unique element in $(\mathcal{S})_H^*$ such that*

$$<< \int_\mathbb{R} Z(t) dt, \psi >> = \int_\mathbb{R} << Z(t), \psi >> dt, \quad \text{for all } \psi \in (\mathcal{S})_H. \qquad (3.4)$$

*Remark*: $t \to B_t^H$ is differentiable in $(\mathcal{S})_H^*$, i.e. fractional noise is a well-defined object and we denote it by $(W_t^H)$.

Below we describe the Wick product which is the last ingredient necessary for describing the integration w.r.t fBm.

DEFINITION 3.2. *Suppose $F, G \in (\mathcal{S})_H^*$ are given by*

$$F(w) = \sum_{\alpha \in \mathcal{J}} a_\alpha H_\alpha(w) \quad \text{and} \quad G(w) = \sum_{\beta \in \mathcal{J}} b_\beta H_\beta(w). \qquad (3.5)$$

*Then the Wick product $F \diamond G$ of $F$ and $G$ is defined as*

$$(F \diamond G)(w) = \sum_{\alpha, \beta \in \mathcal{J}} a_\alpha b_\beta H_{\alpha+\beta}(w). \qquad (3.6)$$

*Remark*: $(\mathcal{S})_H$, and $(\mathcal{S})_H^*$ are closed under Wick product.

The Wick exponential $\exp^\diamond$ is defined as $\exp^\diamond(X) = \sum_{n=0}^\infty \frac{X^{\diamond n}}{n!}$, provided the series converges in $(\mathcal{S})_H^*$, where $X^{\diamond n} = X \diamond \ldots \diamond X$ ($n$ factors). And we have that $\exp^\diamond(<w, f>) = \exp\left(<w, f> - \frac{1}{2}|f|_\phi^2\right)$, for $f \in L_\phi^2(\mathbb{R})$ (see [17]).

DEFINITION 3.3. *Suppose $Y : \mathbb{R} \to (\mathcal{S})_H^*$ is such that $Y(t) \diamond W_t^H$ is integrable in $(\mathcal{S})_H^*$. Then $\int_\mathbb{R} Y(t) dB_t^H$ is defined by*

$$\int_\mathbb{R} Y(t) dB_t^H := \int_\mathbb{R} Y(t) \diamond W_t^H dt. \qquad (3.7)$$

LEMMA 3.4. *Let $\mathcal{L}_\phi^{1,2}(\mu)$ be as in (2.9). If $Y \in \mathcal{L}_\phi^{1,2}(\mu)$, then $\int_\mathbb{R} Y_t dB_t^H$ exists as an element of $L^2(\mu)$ and its norm is given by $\| Y \|_{\mathcal{L}_\phi^{1,2}(\mu)}$.*

For finding the equilibrium strategies we also make use of the Hida derivative (which is called the Malliavin derivative in the context of Wiener space) which we will define below. We first define the directional derivative:

DEFINITION 3.5. *Suppose that $F : \mathcal{S}' \to \mathbb{R}$ and $\gamma \in \mathcal{S}'$. Then the directional (Gateaux) derivative of $F$ in the direction of $\gamma$ is given by*

$$D_\gamma F(w) := \lim_{\epsilon \to 0} \frac{F(w + \epsilon \gamma) - F(w)}{\epsilon}, \quad \text{if it exists in } (\mathcal{S})_H^*. \qquad (3.8)$$

DEFINITION 3.6. $F : \mathcal{S}' \to \mathbb{R}$ *is said to be differentiable if there is a map* $K : \mathbb{R} \to (\mathcal{S})_H^*$ *such that*

$$\begin{aligned} & K(t,w)\gamma(t) \text{ is } (\mathcal{S})_H^* \text{ integrable} \\ & \text{and} \quad D_\gamma F(w) = \int_{\mathbb{R}} K(t,w)\gamma(t)dt \text{ for all } \gamma \in L^2(\mathbb{R}). \end{aligned} \quad (3.9)$$

*Then* $D_t F(w) := K(t,w)$ *is said to be the Hida derivative of* $F$.
We will make use of the Pothoff-Timpel test functions and distributions (see [31] for the definitions of these objects) to define quasi-conditional expectation in the following sections. We denote these spaces by $\mathcal{G}$ and $\mathcal{G}^*$ respectively. The Hida derivative of the random variables in $\mathcal{G}^*$ exist.

Let $F = \sum_\alpha c_\alpha H_\alpha(w) \in \mathcal{G}^*$. Then the Hida derivative exists and is given by

$$D_t F(w) = \sum_\alpha c_\alpha \sum_i \alpha_i H_{\alpha - \varepsilon^i}(w) e_i(t), \quad (3.10)$$

where $\varepsilon^i = (0, ..., 0, 1, 0, ..., 0)$ with the 1 in the $i$th component.

We proceed by defining the quasi-conditional expectation and then introducing the fractional Clark-Ocone theorem which will be crucial in reducing the dynamic optimization problems of the next section into static optimization problems.

DEFINITION 3.7. *[17] If* $F \in \mathcal{G}^*(\mu)$ *has the following expansion*

$$F(w) = \sum_{n=0}^\infty \int_{[0,T]^n} f_n (dB^H)^{\otimes n}, \quad (3.11)$$

*then its quasi-conditional expectation is given by*

$$\tilde{E}_\mu \{F | \mathcal{F}_t\} = \sum_{n=0}^\infty \int_{[0,t]^n} f_n (dB^H)^{\otimes n}. \quad (3.12)$$

Note that $\tilde{E}_\mu \{F | \mathcal{F}_t\} \neq E_\mu \{F | \mathcal{F}_t\}$ in general. (Only for $H = \frac{1}{2}$ is the quasi-conditional expectation operator the same as the conditional expectation operator on $L^2(\mu)$.) However the following holds: $\tilde{E} \{F | \mathcal{F}_t\} = F$ a.s. $\Leftrightarrow F$ is $\mathcal{F}_t$ measurable.

The following feature of the quasi-conditional expectation will be helpful in the computations in the next section:

$$\tilde{E}\{F \diamond G | \mathcal{F}_t\} = \tilde{E}\{F | \mathcal{F}_t\} \diamond \tilde{E}\{G | \mathcal{F}_t\} \quad \text{for } F, G \in \mathcal{G}^*. \quad (3.13)$$

We will also need the notion of a quasi-martingale which is defined as follows:

DEFINITION 3.8. *Suppose* $M_t$ *is an* $(\mathcal{F}_t)$ *adapted process in* $\mathcal{G}^*$. *It is called a quasi-martingale if*

$$\tilde{E}\{M_t | \mathcal{F}_s\} = M_s, \quad \text{for all } t \geq s. \quad (3.14)$$

LEMMA 3.9. *[18] Let* $F \in \mathcal{L}_\phi^{1,2}(\mu)$. *Then* $M_t = \int_0^t F_s dB_s^H$ *is a quasi-martingale.*
Now we can state the fractional Clark-Ocone theorem.

THEOREM 3.10. *[17] Suppose $G(w) \in L^2(\mu)$ is $\mathcal{F}_T$ measurable. Define $\psi(t,w) = \tilde{E}_\mu\{D_t G | \mathcal{F}_t\}$, where $D_t G$ is the Hida derivative of $G$ at $t$, which exists as an element of $\mathcal{G}^*(\mu)$. Then $\psi \in \mathcal{L}_\phi^{1,2}(\mu)$ and*

$$G(w) = E_\mu\{G\} + \int_0^T \psi(t,w) dB_t^H. \tag{3.15}$$

**4. Proof of Theorem 2.2.** Recall that in Theorem 2.2, we consider the one dimensional state equation (2.1) where the pay-off function of player $i$ is of the form (2.2).

As noted previously, we will employ the fractional Clark-Ocone formula and the Wick calculus introduced in Section 3 to find Nash equilibria for this type of game. We begin this development by stating a fractional version of Girsanov's theorem, which is given by [17].

THEOREM 4.1. *([17]) Suppose $T > 0$ and $u : [0,T] \to \mathbb{R}$ is continuous. Suppose further that $K : [0,T] \to \mathbb{R}$ satisfies the equation*

$$\int_0^T K(s)\phi(s,t)ds = u(t), \quad 0 \leq t \leq T, \tag{4.1}$$

*where $\phi$ is given by (2.7). Extend $K$ to $\mathbb{R}$ by putting $K(s) = 0$ outside $[0,T]$. Define the probability measure $\hat{\mu}$ on $F_T$ by*

$$\frac{d\hat{\mu}(w)}{d\mu(w)} = \exp\left(-\int_0^T K(s) dB_s^H - \frac{1}{2}|K|_\phi^2\right). \tag{4.2}$$

*Then $\hat{B}_t^H = \int_0^t u(s)ds + B_t^H$, is an fBm with respect to $\hat{\mu}$.*

The dynamics of the state (2.1) can be written as

$$d(e^{-rt}X_t) - e^{-rt}\sum_{i=1}^N \alpha_i(t)u_i(t)dt = e^{-rt}\sum_{i=1}^N \beta_i(t)v_i(t)(Cdt + dB_t^H). \tag{4.3}$$

Let $\eta$ and $\hat{\mu}$ be defined as in (2.6); i.e.,

$$\begin{aligned}\eta(T) = \frac{d\hat{\mu}}{d\mu} &= \exp\left(-\int_0^T K(s)dB_s^H - \frac{1}{2}|K|_\phi^2\right) \\ &= \exp^\diamond\left(-\int_0^T K(s)dB_s^H\right),\end{aligned} \tag{4.4}$$

where $K$ is from (2.3). Then since $K$ solves (4.1) for $u(t) = C$ (see Appendix Lemma 7.1) and by the fractional Girsanov formula, the process

$$\hat{B}_t^H = Ct + B_t^H, \tag{4.5}$$

is an fBm with respect to $\hat{\mu}$ having the same Hurst parameter as the modulating process in (2.1). Thus, the differential equation describing the flow of the state is given in terms of $\hat{B}^H$ as,

$$e^{-rt}X_t - \int_0^t e^{-rs}\sum_{i=1}^N \alpha_i(s)u_i(s)dt = x + \int_0^t e^{-rs}\sum_{i=1}^N \beta_i(s)v_i(s)d\hat{B}_s^H. \tag{4.6}$$

To be able to find a Nash equilibrium, we will use the quasi-martingale approach to stochastic control in the proof. (For another application of this approach see [18].) We first find the best response of a player to the given strategies of other players, and for that we will use the fractional Clark-Ocone theorem (Thm 3.10).

By (2.8) we have that $e^{-rt}X_t - \int_0^t e^{-rs} \sum_{i=1}^N \alpha_i(s)u_i(s)dt \in L^2(\hat{\mu})$. And note that by Lemma 3.9 and (2.8), $\int_0^t e^{-rs} \sum_{i=1}^N \beta_i(s)v_i(s)d\hat{B}_s^H$, is a quasi-martingale. Therefore we have

$$E_{\hat{\mu}}\left\{e^{-rt}X_t - \int_0^t e^{-rs} \sum_{i=1}^N \alpha_i(s)u_i(s)dt\right\} = x.$$

Now let $G$ be given by

$$G = e^{-rT}F_i - \int_0^T e^{-rs} \sum_{j=1}^N \alpha_j(s)u_j(s)ds. \tag{4.7}$$

Assume $G \in L^2(\hat{\mu})$. Then if

$$E_{\hat{\mu}}\{G\} = x, \tag{4.8}$$

by the fractional Clark-Ocone formula (3.15) we have

$$G = x + \int_0^T \tilde{E}_{\hat{\mu}}\left\{D_s G | \mathcal{F}_s\right\} d\hat{B}_s^H. \tag{4.9}$$

If we choose $v_i$ in (4.6) such that

$$v_i(s) = \frac{-\sum_{j \neq i} \beta_j(s)v_j(s) + e^{rs}\tilde{E}_{\hat{\mu}}\{D_s G | \mathcal{F}_s\}}{\beta_i(s)}, \tag{4.10}$$

then from (4.9) we see that $X_T = F_i$.

By the above argument we can change the dynamic optimization problem of maximizing (2.2) under the dynamics (2.1) into a static optimization problem. In particular, given the other players' strategies, player $i$ wishes to solve the following maximization problem:

$$K_i(x) = \sup_{u_i, F_i}\left\{E_\mu\left\{\int_0^T \frac{c_i u_i(t)^{\gamma_i}}{\gamma_i}dt + \frac{b_i F_i^{\gamma_i'}}{\gamma_i'}\right\}\right\}; \text{ given that}$$

$$E_{\hat{\mu}}\left\{-\int_0^T e^{-rs} \sum_{j=1}^N \alpha_j(s)u_j(s)ds + e^{-rT}F_i\right\} = x\right\}, \tag{4.11}$$

where the supremum is taken over $F_i$ and $(u_i)$ such that

$$e^{-rT}F_i - \int_0^T e^{-rs} \sum_{j=1}^N \alpha_j(s)u_j(s)ds \in L^2(\hat{\mu}). \tag{4.12}$$

This optimization problem can be solved by first considering for each $\lambda_i > 0$ the following unconstrained problem,

$$C_i(x, \lambda) = \sup_{u_i, F_i}\left\{E_\mu\left\{\int_0^T \frac{c_i u_i(t)^{\gamma_i}}{\gamma_i}dt + \frac{b_i F_i^{\gamma_i'}}{\gamma_i'}\right\} + \lambda_i E_{\hat{\mu}}\left\{-\int_0^T e^{-rs} \sum_{j=1}^N \alpha_j(s)u_j(s)ds + e^{-rT}F_i\right\}\right\}, \tag{4.13}$$

and then solving for $\lambda_i$ from the slackness condition:

$$E_{\hat{\mu}}\left\{-\int_0^T e^{-rs}\sum_{j=1}^N \alpha_j(s)u_j(s)ds + e^{-rT}F_i\right\} = x. \quad (4.14)$$

Let us define, as before, the following random variable

$$\rho(t,w) = E_\mu\{\eta(T)|\mathcal{F}_t\}, \quad (4.15)$$

where $\eta$ is from (4.4). Using the fact that

$$E_\mu\{\eta(T)u_i(t)\} = E_\mu\{\rho(t)u_i(t)\}, \quad (4.16)$$

we can solve (4.13) by maximizing pointwise, i.e. for each $t$ and $w$, the functions,

$$g_i(u_i) = \frac{c_i u_i^{\gamma_i}}{\gamma_i} - \lambda_i \rho(t,w)e^{-rt}\sum_{j=1}^N \alpha_j(t)u_j, \quad (4.17)$$

$$\text{and} \quad h_i(F_i) = \frac{b_i F_i^{\gamma_i'}}{\gamma_i'} - \lambda_i \eta(T,w)e^{-rT}F_i. \quad (4.18)$$

Since $0 < \gamma_i < 1$, these functions are concave, and therefore we can solve $g_i'(u_i) = 0$ and $h_i'(F_i) = 0$ to find the maximizing points, which are given by,

$$u_i(t) = \left(\frac{\lambda_i \rho(t,w)e^{-rt}\alpha_i(t)}{c_i}\right)^{\frac{1}{\gamma_i-1}}, \quad (4.19)$$

$$\text{and} \quad F_i = \left(\frac{\lambda_i \eta(T,w)e^{-rT}}{b_i}\right)^{\frac{1}{\gamma_i'-1}}. \quad (4.20)$$

Since $\alpha_i(t)$ is bounded by assumption, (4.12) is satisfied. Note that at the Nash equilibrium $F_i$ is independent of the player index $i$, i.e. $F_i = F^e$ for all $i$. We will use this condition to show that the Lagrange multipliers at the equilibrium are necessarily linear in $b_i$ and then use the slackness condition to actually find their values. First we will find $E_\mu\left\{\eta(T)^{\frac{1}{\gamma_i'-1}}\right\}$. Note that

$$\eta(T)^{\frac{1}{\gamma_i'-1}} = \exp\left(\frac{1}{1-\gamma_i'}\int_0^T K(s)dB_s^H + \frac{1}{2(1-\gamma_i')}|K|_\phi^2\right)$$

$$= \exp\left(\frac{1}{1-\gamma_i'}\int_0^T K(s)dB_s^H - \frac{1}{2(1-\gamma_i')^2}|K|_\phi^2\right)\exp\left(\frac{2-\gamma_i'}{2(1-\gamma_i')^2}|K|_\phi^2\right). \quad (4.21)$$

$$\text{Since} \quad E\left\{\exp^\diamond\left(\int_0^T f(s)dB_s^H\right)\right\} = 1, \quad (4.22)$$

for all $f \in L^2(\mu)$, we have

$$E\left\{\eta(T)^{\frac{1}{\gamma'_i-1}}\right\} = \exp\left(\frac{2-\gamma'_i}{2(1-\gamma'_i)^2}|K|^2_\phi\right). \tag{4.23}$$

Therefore using (4.20) we obtain

$$EF_i = \left(\frac{\lambda_i}{b_i}\right)^{\frac{1}{\gamma'_i-1}} \exp\left(\frac{2-\gamma'_i}{2(1-\gamma'_i)^2}|K|^2_\phi - \frac{rT}{\gamma'_i-1}\right). \tag{4.24}$$

It follows that

$$F_i = E\{F_i\}\exp^\diamond\left(\frac{1}{1-\gamma'_i}\int_0^T K(s)dB_s^H\right). \tag{4.25}$$

From (4.25) we see that for a Nash-equilibrium to exist we necessarily have $\gamma'_i = \gamma'$, and $\lambda^e_i = mb_i$. From (4) we see that $m$ is to be found from the slackness condition:

$$E_\mu\left\{\int_0^T e^{-rt}\rho(t)\left(\sum_{i=1}^N \alpha_i(t)\left(\frac{\lambda^e_i\rho(t)e^{-rt}\alpha_i(t)}{c_i}\right)^{\frac{1}{\gamma_i-1}}\right)dt + e^{-rT}\eta(T)\left(\frac{\lambda^e_i\eta(T)e^{-rT}}{b_i}\right)^{\frac{1}{\gamma'_i-1}}\right\} = x. \tag{4.26}$$

Note that by (4.4) we have the following

$$\begin{aligned}\eta(T)^{\frac{\gamma'}{\gamma'-1}} &= \exp\left(\frac{\gamma'}{1-\gamma'}\int_0^T K(s)dB_s^H + \frac{\gamma'}{2(1-\gamma')}|K|^2_\phi\right) \\ &= \exp^\diamond\left(\frac{\gamma'}{1-\gamma'}\int_0^T K(s)dB_s^H\right)\exp\left(\frac{\gamma'}{2(1-\gamma')^2}|K|^2_\phi\right).\end{aligned} \tag{4.27}$$

$$E\left\{\eta(T)^{\frac{\gamma'}{\gamma'-1}}\right\} = \exp\left(\frac{\gamma'}{2(\gamma'-1)^2}|K|^2_\phi\right). \tag{4.28}$$

Using Thm. 3.2 of [16] $\rho(t,w)$ can be written as

$$\rho(t,w) = \exp\left(-\int_0^t \zeta_t(s)dB_s^H - \frac{1}{2}|\zeta_t|^2_\phi\right), \tag{4.29}$$

where $\zeta_t$ is given by the following:

$$\begin{aligned}((-\Delta)^{-(H-1/2)}\zeta_t)(s) &= ((-\Delta)^{-(H-1/2)}K)(s), \quad 0 \le s \le t, \\ \zeta_t(s) &= 0 \quad s < 0 \quad \text{or} \quad s > t,\end{aligned} \tag{4.30}$$

with the operator $(-\Delta)^{-(H-1/2)}$ on $L^2(\mu)$ defined by (2.5).

Thus

$$E\left\{\rho_t^{\frac{\gamma_i}{\gamma_i-1}}\right\} = E\left\{\exp\left(\frac{\gamma_i}{1-\gamma_i}\int_0^T \zeta_t(s)dB_s^H - \frac{\gamma_i^2}{2(1-\gamma_i)^2}|\zeta_t|^2_\phi + \frac{\gamma_i}{2(1-\gamma_i)}|\zeta_t|^2_\phi + \frac{\gamma_i^2}{2(1-\gamma_i)^2}|\zeta_t|^2_\phi\right)\right\}, \tag{4.31}$$

from which we conclude that

$$E\left\{\rho(t)^{\frac{\gamma_i}{\gamma_i-1}}\right\} = \exp\left(\frac{\gamma_i}{2(1-\gamma_i)^2}|\zeta_t|_\phi\right), \tag{4.32}$$

so that $m$ can be solved from (4.26), which leads to the following equation:

$$\sum_{i=1}^{N} \int_0^T m^{\frac{1}{\gamma_i-1}} b_i^{\frac{1}{\gamma_i-1}} \alpha_i(t)^{\frac{\gamma_i}{\gamma_i-1}} e^{-rt\frac{\gamma_i}{\gamma_i-1}} \exp\left(\frac{\gamma_i}{2(1-\gamma_i)^2}|\zeta_t|_\phi\right) dt \\ + m^{\frac{1}{\gamma'-1}} e^{-rT\frac{\gamma'}{\gamma'-1}} \exp\left(\frac{2\gamma'|K|_\phi}{2(1-\gamma')^2}\right) = x. \quad (4.33)$$

After solving for $m$ using (4.33), then by (4.19) and (4.20) we have the final state at the equilibrium and strategy $u_i$ for player $i$ leading to that state, given respectively by,

$$F^e = m^{\frac{1}{\gamma'-1}} \eta(T)^{\frac{1}{\gamma'-1}} e^{\frac{-rT}{\gamma'-1}}, \quad (4.34)$$

and

$$u_i^e(t) = \left(\frac{mb_i\alpha_i(t)}{c_i} e^{-rt} \rho(t,w)\right)^{\frac{1}{\gamma_i-1}}. \quad (4.35)$$

Observe that these controls are not Markovian. (In a Brownian motion setting the controls were assumed to be Markovian at the outset so that the HJB equations for an equilibrium solution can be developed [7], [12] and [29].)

Now we will proceed to find $(v_i)$ at the equilibrium, which is the second component of the players' strategies. For this we will again make use of the fractional Clark-Ocone formula.

Suppose $G^e$ is given by

$$G^e = e^{-rT} F^e - \int_0^T e^{-rs} \sum_{i=1}^N \alpha_i u_i^e(s) ds. \quad (4.36)$$

Since there is a unique adapted process $\psi(t,w)$ such that

$$G^e = E_\mu\{G^e\} + \int_0^T \psi(t,w) d\hat{B}_t^H, \quad (4.37)$$

which, from the Clark-Ocone formula, is given by

$$\psi(t,w) = \tilde{E}_{\hat{\mu}}\{D_t G^e | \mathcal{F}_t\}, \quad (4.38)$$

it can now be seen immediately that any adapted $(v_i^e)$ that satisfies

$$\tilde{E}_\mu\{D_t G^e | \mathcal{F}_t\} = e^{-rt} \sum_{i=1}^N \beta_i v_i^e(t), \quad (4.39)$$

is an equilibrium strategy.

To obtain a more explicit expression, we will compute $\tilde{E}_{\hat{\mu}}\{D_t G^e|\mathcal{F}_t\}$. Using (4.34) and (4.35), $G^e$ is given by

$$G^e(T,w) = m^{\frac{1}{\gamma'-1}} e^{-rT\frac{\gamma'}{\gamma'-1}} \eta(T,w)^{\frac{1}{\gamma'-1}} - \int_0^T \sum_{i=1}^N \alpha_i(t)^{\frac{\gamma_i}{\gamma_i-1}} \left(\frac{mb_i}{c_i}\right)^{\frac{1}{\gamma_i-1}} e^{-rt\frac{\gamma_i}{\gamma_i-1}} \rho(t,w)^{\frac{1}{\gamma_i-1}} dt. \quad (4.40)$$

To calculate the quasi-conditional expectation of the Hida derivative of $G^e$ we will first find it for the stochastic part of the first term on the right-hand side of (4.40). Define $R$ as

$$R = \exp\left(\frac{2-\gamma'}{(1-\gamma')^2}|K|_\phi^2 - \frac{C}{1-\gamma'}\int_0^T K(s)ds\right). \tag{4.41}$$

Using the chain rule, (4.5) and (3.12), we have

$$\begin{aligned}
\tilde{E}_{\hat{\mu}}\left\{D_t\eta(T)^{\frac{1}{\gamma'-1}}|\mathcal{F}_t\right\} &= \tilde{E}_{\hat{\mu}}\left\{\frac{K(t)}{1-\gamma'}\eta(T)^{\frac{1}{\gamma'-1}}|\mathcal{F}_t\right\} \\
&= \frac{K(t)}{1-\gamma'}R\tilde{E}_{\hat{\mu}}\left\{\exp^\diamond\left(\frac{1}{1-\gamma'}\int_0^T K(s)d\hat{B}_s^H\right)|\mathcal{F}_t\right\} \\
&= \frac{K(t)}{1-\gamma'}R\exp^\diamond\left(\frac{1}{1-\gamma'}\int_0^t K(s)d\hat{B}_s^H\right) \\
&= \frac{K(t)}{1-\gamma'}\exp\left(\frac{1}{1-\gamma'}\int_0^T K(s)dB_s^H - \frac{C}{1-\gamma'}\int_t^T K(s)ds\right. \\
&\quad \left.+ \frac{2-\gamma}{2(1-\gamma^2)}|K|_\phi^2 - \frac{1}{1-\gamma}|K1_{[0,t]}|_\phi^2\right).
\end{aligned} \tag{4.42}$$

Now we will find the quasi-conditional expectation of the Hida derivative of the stochastic part of second term on the right-hand side of (4.40) using (4.29). I.e.,

$$\begin{aligned}
\tilde{E}_{\hat{\mu}}\left\{D_t\left(e^{-ru\frac{\gamma_i}{\gamma_i-1}}\rho(u)^{\frac{1}{\gamma_i-1}}\right)|\mathcal{F}_t\right\} &= \\
\frac{\zeta_u(t)}{1-\gamma_i}e^{-ru\frac{\gamma_i}{\gamma_i-1}}&\exp\left(\frac{1}{1-\gamma_i}\int_0^t \zeta_u(s)dB_s^H - \frac{C}{1-\gamma_i}\int_t^u \zeta_u(s)ds\right. \\
&\quad \left.+ \frac{2-\gamma_i}{2(1-\gamma_i)^2}|\zeta_u|_\phi^2 - \frac{1}{1-\gamma_i}|\zeta_u 1_{[0,t]}|_\phi^2\right).
\end{aligned} \tag{4.43}$$

Using (4.39) and (4.40) we have the result for the second component for the players' equilibrium strategies, and this concludes the proof of Theorem 2.2. $\square$

**5. Extension of the Wick Calculus to Arbitrary Gaussian Processes.** Although the results of the preceding sections have considered the explicit case in which the modulator in (1.1) is fBm, these results can be extended to the situation in which the modulator is a more general Gaussian process within sufficient regularity. This requires the extension of the Wick calculus to more general Gaussian processes. In this section, we sketch how this extension can be accomplished. The first step in extending the fractional noise machinery introduced in Section 3 to more general Gaussian processes is the following theorem due to Loève [23] for integrating deterministic functions with respect to second order processes:

THEOREM 5.1. *([23]) Suppose that $X$ is a zero-mean process such that $E\{X_t^2\} < \infty$ for all $t$, and denote its covariance function by $R$. Then, for $-\infty < a < b < \infty$,*

$$\int_a^b f(t)dX_t \tag{5.1}$$

*exists as the $L^2$-limit of Riemann sums if and only if*

$$|f|_R^2 := \int_a^b \int_a^b f(t)f(s)d^2R(s,t) < \infty. \tag{5.2}$$

Henceforth $X$ will denote a Gaussian process. By the Bochner-Minlos theorem, there exists a unique probability measure on the space of tempered distributions such that $<\cdot,f>: \Omega \to \mathbb{R}$ is a Gaussian random variable with mean 0, and variance $|f|_R^2$.

We will denote $L^2(\mu)$ by $L^2(X)$ and $H(X)$ will denote the linear space of $X$, *i.e.* the closed subspace of $L^2(X)$ spanned by $X_t$ for all $t \in [a,b]$ (*i.e.* the first Wiener chaos). As in [19] we construct $\Lambda(R)$, a Hilbert space of deterministic integrable 'functions' isomorphic to $H(X)$ by completing the pre-Hilbert space of step functions $\mathbb{S}$ with the following inner product,

$$<f,g>_\mathbb{S} = \int \int f(t)g(s)d^2R(t,s), \tag{5.3}$$

for any $f, g \in \mathbb{S}$. Then the integration operator defined on the set of step functions (the integration with respect to $X$) can be extended to an isomorphism between $H(X)$ and $\Lambda(R)$. (The elements of $\Lambda(R)$ are generalized functions, *i.e.* distributions [30].)

As a second step we will define the Wick-integrability of a random process with respect to a Gaussian process. This is done by using the tensor product structure of the space $L^2(X)$. Let us define the tensor product of Hilbert spaces.

DEFINITION 5.2. *The algebraic tensor product $H_1 \otimes H_2$ of Hilbert spaces $H_1$ and $H_2$ is a pre-Hilbert space with the following inner product*

$$<h_1 \otimes h_2, g_1 \otimes g_2>_{H_1 \otimes H_2} := <h_1, g_1>_{H_1} <h_2, g_2>_{H_2}, \tag{5.4}$$

*for $g_i, h_i \in H_i$ and $i = 1, 2$. The closure of this pre-Hilbert space is the tensor product of Hilbert spaces, which will still be denoted by $H_1 \otimes H_2$. $H_1 \tilde{\otimes} H_2$ will denote the symmetrized tensor product.*

Then we have the following Wiener chaos isomorphism theorem.

THEOREM 5.3. *([21]) $\oplus_{p \geq 0} H^{\tilde{\otimes}p}(X)$ is isomorphic to $L^2(X)$ with the unique isomorphism $\Phi$ defined by*

$$\Phi(\xi_1^{\tilde{\otimes}\alpha_1} \tilde{\otimes} ... \tilde{\otimes} \xi_k^{\tilde{\otimes}\alpha_k}) = \frac{1}{\sqrt{p!}} \prod_{j=1}^k h_{\alpha_j}(\xi_j), \tag{5.5}$$

*where $\xi_i \in H(X)$ for all $i$ are orthonormal; $p = |\alpha| = \alpha_1 + ... + \alpha_k$. Here $h_n$ is the Hermite polynomial of degree $n$ (see [15], Appendix C ).*

Note that for a random variable $\xi \in H(X)$ with unit variance we have

$$\Phi(e^{\tilde{\otimes}\xi}) = \exp\left(\xi - \frac{1}{2}\right), \tag{5.6}$$

where the exponential is defined by $e^{\tilde{\otimes}\xi} = \sum_{p \geq 0} \frac{\xi^{\tilde{\otimes}p}}{\sqrt{p!}}$.

We proceed as in [19], and in order to define the integral of a stochastic process with respect to $X$, we first define a tensor product integral, denoted by $\int F_t \otimes dX_t$, and its domain, denoted by $\Lambda(R)_{L^2(X)}$.

Suppose $\mathbb{S}_{L^2(X)}$ is the pre-Hilbert space of the $L^2(X)$ valued step functions $F_t$,

$$F_t = \sum_{i=1}^N F_i 1_{(t_i, t_{i+1}]}, \tag{5.7}$$

for $(t_i, t_{i+1}] \in [a,b]$, and $F_i \in L^2(X)$, equipped with the inner product

$$<F,G> = \int\int <F_t, G_s>_{L^2(X)} d^2 R(t,s). \tag{5.8}$$

Let $\Lambda(R)_{L^2(X)}$ denote the completion of $\mathbb{S}_{L^2(X)}$. For the $F \in \mathbb{S}_{L^2(X)}$ given in (5.7) define the integral $I_\otimes$ as

$$\int F_t \otimes dX_t = \sum_{i=1}^{N} F_{t_i} \otimes (X_{t_{i+1}} - X_{t_i}). \tag{5.9}$$

Since this integral is a norm preserving linear map, it has a unique extension to an isomorphism from $\Lambda(R)_{L^2(X)}$ into $L^2(X) \otimes H(X)$. We will construct a map $\Psi$ from $L^2(X) \otimes H(X)$ into $L^2(X)$ and call the composition of the two maps, $\Psi(I_\otimes)$, the stochastic integral. We start by defining the following linear map

$$\Psi_p : H^{\tilde{\otimes}p}(X) \otimes H(X) \to H^{\tilde{\otimes}p+1}(X) \tag{5.10}$$

by

$$\Psi_p\left(\left(\xi_1^{\tilde{\otimes}\alpha_1}\tilde{\otimes}...\tilde{\otimes}\xi_k^{\tilde{\otimes}\alpha_k}\right) \otimes \xi_l\right) = (p+1)^{\frac{1}{2}} \xi_1^{\tilde{\otimes}\alpha_1}\tilde{\otimes}...\tilde{\otimes}\xi_k^{\tilde{\otimes}\alpha_k}\tilde{\otimes}\xi_l, \tag{5.11}$$

where $(\xi_i) \in H(X)$ is an orthonormal set of random variables, and $\alpha_1 + ... + \alpha_k = p$. $\Psi_p$ can be extended uniquely to a bounded linear map with norm $(p+1)^{1/2}$ from $H^{\tilde{\otimes}p}(X) \otimes H(X)$ onto $H^{\tilde{\otimes}p+1}(X)$.

Now define $\Psi^*$ as the map from $\oplus_{p\geq 0} H^{\tilde{\otimes}p}(X) \otimes H(x)$ onto $\oplus_{p\geq 1} H^{\tilde{\otimes}p}(X)$; by $\Psi^* = \oplus_{p\geq 0} \Psi_p$, viz. the restriction of $\Psi^*$ to $H^{\tilde{\otimes}p}(X) \otimes H(X)$ is $\Psi_p$. The domain of the operator $\Psi^*$ is given by

$$\mathcal{D}^* = \left\{\eta \in \left(H^{\tilde{\otimes}\alpha_1}(X) \oplus ... \oplus H^{\tilde{\otimes}\alpha_m}(X)\right) \otimes H(X) : \alpha_1 + ... + \alpha_m < \infty\right\}, \tag{5.12}$$

so that $\sum_{p\geq 0} \|\Psi_p(\eta_p)\|^2 < \infty$, where $\eta_p$ is the projection of $\eta$ on $H^{\tilde{\otimes}p}(X) \otimes H(X)$.

By Thm. 5.3, $\oplus_{p\geq 0} H^{\tilde{\otimes}p}(X)$ is isomorphic to $L^2(X)$. Therefore $\left(\oplus_{p\geq 0} H^{\tilde{\otimes}p}(X)\right) \otimes H(X)$ is isomorphic to $L^2(X) \otimes H(X)$. Denote this isomorphism by $\Phi_0$. Let $\mathcal{D} = \Phi_0(\mathcal{D}^*)$, which is a proper subset of $L^2(X) \otimes H(X)$. Then define $\Psi$ by

$$\Psi = \Phi \circ \Psi^* \circ \Phi_0^{-1}. \tag{5.13}$$

We define the Wick product of $V \in L^2(X)$ and $W \in H(X)$ as

$$V \diamond W := \Psi(V \otimes W). \tag{5.14}$$

Note that $V \diamond W$ is in $L^2(X)$ iff $V \otimes W \in \mathcal{D} \otimes H(X)$.

The integral $\int F_t \diamond dX_t$ is then defined by

$$\int_a^b F_t \diamond dX_t = \Psi \circ I_\otimes(F) \tag{5.15}$$

for all $F$ such that $I_\otimes(F) = \int F_t \otimes dX_t \in \mathcal{D}$. The set of all $F$'s in the domain of integration is denoted by $\Lambda(R)^*_{L^2(X)}$. Then we have the Itô representation formula as a result of the Multiple Wiener Integral (MWI) representation of the random variables in $L^2(X)$ ([19]), and the fact that each MWI can be written as an iterated integral:

THEOREM 5.4. *([19]) Every $\theta \in L^2(X)$ has the following representation*

$$\theta = E\{\theta\} + \int_a^b F_t \diamond dX_t, \qquad (5.16)$$

for an $F \in \Lambda(R)^*_{L^2(X)}$ that is adapted to the filtration generated by $X$.

Now let us define the Wick product of two elements in $L^2(X)$. As a first step we define $\Upsilon_{p,q}$,

$$\Upsilon_{p,q} : H^{\tilde{\otimes}p}(X) \otimes H^{\tilde{\otimes}q}(X) \to H^{\tilde{\otimes}p+q}(X), \qquad (5.17)$$

as

$$\Upsilon_{p,q}\left(\left(\xi_{\gamma_1}^{\tilde{\otimes}\alpha_1}\tilde{\otimes}...\tilde{\otimes}\xi_{\gamma_k}^{\tilde{\otimes}\alpha_k}\right) \otimes \left(\xi_{\lambda_1}^{\tilde{\otimes}\beta_1}\tilde{\otimes}...\tilde{\otimes}\xi_{\lambda_l}^{\tilde{\otimes}\beta_l}\right)\right) = \sqrt{\frac{(p+q)!}{p!q!}}\xi_{\gamma_1}^{\tilde{\otimes}\alpha_1}\tilde{\otimes}...\tilde{\otimes}\xi_{\gamma_k}^{\tilde{\otimes}\alpha_k}\tilde{\otimes}\xi_{\lambda_1}^{\tilde{\otimes}\beta_1}\tilde{\otimes}...\tilde{\otimes}\xi_{\lambda_l}^{\tilde{\otimes}\beta_l}, \qquad (5.18)$$

for any $(\xi_\gamma)$ that is an orthonormal set in $H(X)$.

On denoting $\Upsilon = \oplus_{p \geq 0} \oplus_{q \geq 0} \Upsilon_{p,q}$, we define the Wick-product of the $W, V \in L^2(X)$ as

$$W \diamond V := \Phi\left(\Upsilon\left(\Phi^{-1}(W) \otimes \Phi^{-1}(V)\right)\right). \qquad (5.19)$$

Note that $L^2(X)$ is not closed under $\diamond$, since the tensor product of the random variables may not be in the domain of $\Upsilon$. Then one can define the Hida distribution space, use (5.19) as the definition of the Wick-product over this space, and see that the Wick product so defined is closed over these spaces.

The main machinery we use to develop strategies leading to a Nash equilibrium are the Girsanov formula (the absolute continuity of the translated measure w.r.t. the original measure), and the Clark-Ocone formula. These can be extended to more general Gaussian modulators with sufficient regularity. The Girsanov theorem, Thm. 4.1, can be stated for an sufficiently regular Gaussian processes. (The proof of the Girsanov theorem in [17] does not make use of the explicit expression for $\phi$.) The derivation of the Clark-Ocone theorem (Thm. 3.10) is done by using only the tensor product structure of the space $L^2$ (Thm. 5.3) and the spaces of generalized random variables. Defining the Hida derivative and the quasi-conditional expectation operator (w.r.t. which $X$ is a quasi-martingale) for the Gaussian process $X$, we can restate the Clark-Ocone theorem. Now replacing $\zeta_t$ in Thm. 2.2 by $\vartheta_t$ such that

$$E\left\{\int_0^T K(s) dX_s \bigg| \mathcal{F}_t\right\} = \int_0^t \vartheta_t(s) dX_s, \qquad (5.20)$$

we have a Nash equilibrium theorem for a general Gaussian process. Note that, unlike the case of fBm we cannot in general write $\vartheta$ explicitly in terms of $K$. Hence, we cannot give an explicit solution for the Nash equilibrium. A general multi-dimensional theorem can also be restated for a multi-dimensional Gaussian process with independent components (the components do not have to be identical) by making same conceptual modifications as in the one-dimensional case.

**6. Conclusion.** In this paper we have explicitly found Nash equilibria for stochastic differential games in a non-Markovian setting. In this formalism, all the agents observe the states, and they control the states by modifying the drift and the volatility. The agents are heterogeneous in their controls and utility functions. We have taken the modulating process to be fractional Brownian motion, because an fBm is versatile in modeling long-range dependence phenomena in finance and networks.

Since the diffusion in our model is modulated by a non-Markovian process, the usual technique of finding Nash equilibria via Hamilton-Jacobi-Bellman equations is not available. Therefore we have made use of the fractional noise calculus to calculate the agents' Nash-equilibrium strategies. Although we have taken the modulating process of the diffusions to be fBm, our results hold for more general Gaussian modulating processes with only slight modifications to the white noise machinery.

Our results are applicable to financial markets in which stock price dynamics are modulated with fractional Brownian motion. One of the candidate applications is stock price modeling when each agent's activities in the market affect the price flow (institutional investors are such examples), or if there are transaction costs. This work is also applicable to stochastic portfolio games, in which agents compete for a bonus.

**7. Appendix.** LEMMA 7.1. *([22]) Let $f : [0, T] \to \mathbb{R}$ be a continuous function and introduce the following integral equation:*

$$\int_0^T \hat{f}(s)\phi(s,t)ds = f(t) \quad \text{for } t \in [0,T], \tag{7.1}$$

*where $\phi$ is given by (2.7). The solution to this equation is given by*

$$\hat{f}(t) = -\frac{1}{d_H} t^{\frac{1}{2}-H} \frac{d}{ds} \int_t^T dw \, w^{2H-1}(w-t)^{\frac{1}{2}-H} \frac{d}{dw} \int_0^w dz \, z^{\frac{1}{2}-H}(w-z)^{\frac{1}{2}-H} f(z), \tag{7.2}$$

*where*

$$d_H = 2H(2H-1)\left(\Gamma(\frac{3}{2}-H)\right)^2 \Gamma(2H-1)\cos(\pi(H-\frac{1}{2})) \tag{7.3}$$

COROLLARY 7.2. *If we take $f(t)=C$ on $[0,T]$ in the integral equation given by (7.1) then the solution $\hat{f}(t)$ is given by*

$$\hat{f}(t) = \frac{C}{k_H} t^{\frac{1}{2}-H}(T-t)^{\frac{1}{2}-H}, \tag{7.4}$$

*where*

$$k_H = 2H(2H-1)\Gamma(2-2H)\Gamma(2H-1)\cos(\pi(H-\frac{1}{2})). \tag{7.5}$$

*Proof:* The proof can be found in [17], but we present it here for the sake of completeness.

$$\hat{f}(t) = -\frac{1}{d_H} C t^{\frac{1}{2}-H} \frac{d}{ds} \int_t^T dw \, w^{2H-1}(w-t)^{\frac{1}{2}-H} \frac{d}{dw} \int_0^w dz \, z^{\frac{1}{2}-H}(w-z)^{\frac{1}{2}-H}. \tag{7.6}$$

Note that

$$\frac{\int_0^w z^{\frac{1}{2}-H}(w-z)^{\frac{1}{2}-H} dz}{w^{2-2H}} = B(\frac{3}{2},\frac{3}{2}) = \frac{\Gamma\left(\frac{3}{2}-H\right)^2}{\Gamma(3-H)}, \tag{7.7}$$

where $B(\cdot,\cdot)$ is the beta function given by

$$B(x,y) = \int_0^1 t^{x-1}(1-t)^{y-1}dt. \tag{7.8}$$

Hence

$$\frac{d}{dw}\int_0^w z^{\frac{1}{2}-H}(w-z)^{\frac{1}{2}-H}dz = \frac{\Gamma\left(\frac{3}{2}-H\right)^2}{\Gamma(2-H)}w^{1-2H}. \tag{7.9}$$

Using (7.9) it is not hard to evaluate (7.6) to get (7.4). $\square$